\documentclass[preprint,11pt]{elsarticle}   

\usepackage{amssymb}
\usepackage{amsmath}
\usepackage{amsthm}
\usepackage{graphicx}
\usepackage{float}
\usepackage{booktabs}
\usepackage{lineno}

\usepackage{geometry}

\journal{Physics of the Dark Universe}

\begin{document}

\begin{frontmatter}
\vspace{-3cm}
\title{\boldmath \huge Alleviating the Hubble Tension Using Coupled Dark Energy - Dark Matter Interaction}

\author[a,b]{Yismaw Wassie\corref{cor1}} 
\ead{yismawwassie2015@gmail.com}
\author[c,d]{Amare Abebe}        
\author[a,e]{Solomon Belay}    
\author[c,d]{Shambel Sahlu}  

\cortext[cor1]{Corresponding author.}

\affiliation[a]{organization={Department of Astronomy and Astrophysics, Entoto Observatory and Research Center (EORC), Space Science and Geospatial Institute (SSGI)},
            city={Addis Ababa},
            country={Ethiopia}}

\affiliation[b]{organization={Addis Ababa University},
            city={Addis Ababa},
            country={Ethiopia}}

\affiliation[c]{organization={Centre for Space Research, North-West University},
            city={Potchefstroom},
            postcode={2520},
            country={South Africa}}

\affiliation[d]{organization={National Institute for Theoretical and Computational Sciences (NITheCS)},
            city={Potchefstroom},
            postcode={2520},
            country={South Africa}}

\affiliation[e]{organization={Ministry of Innovation and Technology (MinT)},
            city={Addis Ababa},
            country={Ethiopia}}


\begin{abstract}
\noindent We investigate two interacting dark energy models, $\xi$CDM and $\xi w$CDM, as potential means of alleviating the $H_0$ tension, using DESI DR2 BAO data along with the Planck 2018 CMB, Pantheon+ SNe Ia, and KiDS-1000/DES-Y3 weak lensing data. For the $\xi$CDM model, with a free coupling strength ($\xi$) and fixed dark energy equation-of-state ($w_0$), we find $H_0 = 71.59^{+1.69}_{-2.30}\mathrm{kms^{-1}Mpc^{-1}}$ from Planck-only data, tightening to $69.72^{+0.50}_{-0.41}\mathrm{kms^{-1}Mpc^{-1}}$ for the complete Planck+BAO+SN+WL dataset, reducing the tension from $\sim5.0\sigma$ to $\sim2.8\sigma$. In the $\xi w$CDM model, with both $\xi$ and $w_0$ free, Planck-only data provide $H_0 = 72.46^{+4.05}_{-3.66}\mathrm{kms^{-1}Mpc^{-1}}$, while the full dataset yields $H_0 = 70.22^{+0.45}_{-0.40}\mathrm{kms^{-1}Mpc^{-1}}$, reducing the tension to $\sim2.5\sigma$, with phantom preference $w_0 = -1.116^{+0.043}_{-0.043}$ and negative coupling $\xi = -0.107^{+0.012}_{-0.023}$. With the convention $-\xi H \rho_{de}$ >0, the negative coupling ($\xi < 0$) points to $Q > 0$, giving continuous energy transfer from dark energy to dark matter. This interaction modifies the evolution of the dark sector and the expansion history from early to late times, shifting the inferred $H_0$ to higher values. For the $\xi w$CDM model, Planck+BAO+SN yields a substantial improvement in fit ($\Delta\chi^2=-14.60$) and a strong AIC preference ($\Delta\mathrm{AIC}=-10.60$), but $\Delta \mathrm{BIC} = +0.30$ weakly prefers $\Lambda$CDM. Inclusion of WL data weakens the fit improvement ($\Delta\chi^2 = -6.67$) yielding $\Delta \mathrm{BIC} = +8.23$, a strong BIC preference for $\Lambda$CDM. This AIC/BIC discrepancy shows that the data favor the extra dark sector freedom, but not strongly enough to overcome the BIC penalty. These interacting models are therefore viable mechanisms for alleviating the $H_0$ tension.
\end{abstract}




\end{frontmatter}



\section{Introduction}
\label{sec:intro}
\noindent The standard cosmological model, $\Lambda$CDM, has emerged as the standard framework in modern cosmology, successfully describing a broad range of observations including cosmic microwave background (CMB) anisotropies~\cite{Planck2020}, the large scale distribution of galaxies~\cite{Alam2017}, and the accelerated expansion of the universe inferred from Type Ia supernovae~\cite{Riess1998,Perlmutter1999}. However, this model faces persistent and growing challenges that have intensified with the increasing precision of cosmological observations. Among these,  the discrepancy between early and late universe measurements of the Hubble constant $H_0$ stands as one of the most significant anomalies in contemporary cosmology~\cite{Riess1998,Perlmutter1999,Verde2019,DiValentino2021, Verde2019,Abdalla2022,Kamionkowski2023}. Within the $\Lambda$CDM model, Planck Satellite's CMB observations at redshift $z \sim 1100$ yield $H_0 = 67.36 \pm 0.54\mathrm{kms^{-1}Mpc^{-1}}$~\cite{Planck2020}. In contrast, the SH0ES collaboration's  local distance ladder measurement using Type Ia supernovae calibrated by Cepheids presents $H_0 = 73.04 \pm 1.04\mathrm{kms^{-1}Mpc^{-1}}$~\cite{Riess2022}. This tension now greater than $5\sigma$ remains consistent despite significant work to find systematic uncertainties~\cite{George2021,Jang2021}. This  continuing tension implies the necessity of new physics beyond the standard cosmological model or unresolved systematic uncertainties \cite{Abdalla2022, Kamionkowski2023, DiValentino2025}.\\
A wide range of theoretical modifications have been proposed to address the Hubble tension using different mechanisms. These approaches can be classified into two categories: early time modifications, which alter the pre-recombination universe (such as early dark energy~\cite{Poulin2019}), and late-time modifications (post-recombination evolution) involving dark sector interactions and gravitational modifications~\cite{Amendola2000, Sahni2002, Yang2021, Clemson2012}. From these, the non-gravitational interaction between dark energy and dark matter is well motivated~\cite{Amendola2012}. The comparable energy densities of dark matter and dark energy in late time naturally lead to the energy exchange between the two dominant components of the universe ~\cite{Amendola2012, Wang2026, Wang:2016lxa}.\\
The interacting dark energy (IDE) model introduces a phenomenological coupling parameter $\xi$, that controls the coupling strength and direction of the energy transfer between the two dark sectors of the universe. Following the covariant formulation introduced in Ref. \cite{Gavela2009,Gavela2010}, we consider an interaction of the form $Q=\xi H \rho{\rm de}$, where the stability against cosmological perturbation growth requires a condition $\xi(1+w_0)>0$~\cite{Majerotto2009,Jackson2009}. Here, $w_0$ is a parameter that indicates dark energy equation-of-state. For the phantom regime, where $w_0<-1$, the stability condition requires a negative coupling, $\xi<0$, equivalent to energy transfer from dark energy to dark matter under our sign convention. This energy transfer leads to a deviation of the dark-matter density from the standard $a^{-3}$ dilution behavior, which modifies the evolution of dark matter from the early universe onward. Hence, to keep the current value of physical cold dark matter density $\Omega_c h^2$, fixed from the CMB acoustic peaks, the modified evolution of dark matter allows a higher inferred $H_0$, thereby alleviating the $H_0$ tension~\cite{DiValentino2020a, DiValentino2020b}. \\
Various studies have examined IDE models as a way of alleviating the $H_0$ tension, using earlier BAO measurements, with some analysis getting a reduction of the tension to approximately $2.6$--$3\sigma$~\citep{DiValentino2020a,DiValentino2020b, Pan2019, Salvatelli2013,  Rugg:2024}. 
Recently, there have been a limited number of studies employing the newly available DESI DR2 BAO data to examine IDE models~\citep{Silva2025, Zhang2025a, Aboubrahim2026, Hoerning2025, Petri2026}. These studies show  the capability of the DESI DR2 BAO data in providing new constraints on dark-sector interactions. However, a comprehensive analysis considering the DESI DR2 BAO measurements along with Planck CMB, Pantheon+ SNe Ia, and weak-lensing data is still required in order to assess the potential of IDE models in alleviating the $H_0$ tension. In this work, we therefore perform new analysis for two interacting dark energy models using DESI DR2 BAO, Planck 2018 CMB, Pantheon+ SNe, and KiDS-1000/DES-Y3 weak lensing data.  One of the interacting energy models is $\xi$CDM in which $\xi$ is taken to be free, while $w_0$ is fixed at -1.001, phantom surrogate. The other model is $\xi w$CDM model where both parameters, $\xi$ and $w_0$, are allowed to vary. \\
Our work improves on previous studies through the inclusion of precise DESI DR2 BAO, updated pantheon+ SNe, KiDs-1000/DES-Y3 weak lensing data, phantom branches exploration, and detailed AIC/BIC assessments. Within this framework, the $\xi$CDM model provides $H_0 = 71.59^{+1.69}_{-2.30}$ km s$^{-1}$ Mpc$^{-1}$ from Planck-only dataset, tightening to  $69.72^{+0.50}_{-0.41}$ km s$^{-1}$ Mpc$^{-1}$  with the complete dataset combination (Planck+BAO+SN+WL), minimizing the $H_0$ tension from $\sim 5.0\sigma$ to $\sim 2.8\sigma$. The $\xi w$CDM model on the other hand  yields more favorable results with $H_0 = 72.46^{+4.05}_{-3.66}$ km s$^{-1}$ Mpc$^{-1}$ from Planck-only case, $H_0 = 71.10^{+0.83}_{-0.45}$ km s$^{-1}$ Mpc$^{-1}$ for the Planck+BAO+SN, and  $H_0 = 70.22^{+0.45}_{-0.40}$ km s$^{-1}$ Mpc$^{-1}$  for the complete dataset combinations, reducing the $H_0$ to $\sim 2.5\sigma$. For the $\xi w$CDM, the Planck+BAO+SN combination yields $\Delta\chi^2 = -14.60$, showing a clear progress in fit, but the $\Delta BIC = +0.30$ due to parameter penalty, i.e weak preference. The inclusion of WL weakens the fit improvment ($\Delta\chi^2 = -6.67$), so BIC shows a strong preference for $\Lambda CDM$ ($\Delta BIC = +8.23$). AIC, which indicates weaker penalty on additional parameters, supports the interacting model in both cases ($\Delta AIC = -10.60 and -2.67$). This AIC/BIC discrepancy shows that the data favor the extra dark sector freedom, but not with sufficient strength to overcome the penalty under BIC for large N. We thus state the interacting models as physically motivated and empirically viable mechanism for alleviating the $H_0$ tension.\\
This paper is structured as follows. In Section~\ref{subsec:lambdas_model}, we outline the theoretical motivation of the IDE model, that includes background evolution, perturbation equations, and stability criteria. In Section~\ref{sec:methods}, we present the methodology which covers the modification of CLASS code, data, and parameter estimation technique. The main results are discussed in Section~\ref{sec:results}, including parameter constraints, statistical comparison of models, and $H_0$ tension analysis. We conclude with a discussion of our findings and their implications in Section~\ref{sec:conc}.

\section{The IDE Model and Theoretical Motivation}
\label{subsec:lambdas_model}
\noindent The IDE model develops a dark energy-dark matter interaction operating independently of gravitational forces. This interaction is covariantly presented through energy-momentum transfer four-vectors~\cite{Gavela2009,Majerotto2009} with the form:
\begin{align}
    \nabla_\mu T^{\mu\nu}_{\mathrm{(dm)}} &= -Q^\nu\;, \label{eq:cov_dm}\\
	\nabla_\mu T^{\mu\nu}_{\mathrm{(de)}} &= +Q^\nu\;, \label{eq:cov_de} 
\end{align}
with the combined dark sector satisfying $\nabla_\mu(T^{\mu\nu}{\mathrm{(de)}} + T^{\mu\nu}{\mathrm{(dm)}}) = 0$. Following previous works~\cite{Yang2021, Amendola2000a, vanderWesthuizen:2023}, we use the form of interaction given by:
\begin{equation}
	 \quad Q = \xi H \rho_{\mathrm{de}},
	\label{eq:interaction}
\end{equation}
where $H$ represents the Hubble parameter, $\rho_{\mathrm{de}}$ is the dark energy density, and $\xi$ is a constant that refers to interaction strength and direction of energy flow.\\
For a spatially flat FLRW metric, the background continuity equations reduce to ~\cite{Gavela2009}:
\begin{align}
\dot{\rho}_{\rm dm} + 3H\rho_{\rm dm} &= -\xi H \rho_{\rm de}, \label{eq:dm_cont} \\
\dot{\rho}_{\rm de} + 3H(1+w_0)\rho_{\rm de} &= +\xi H \rho_{\rm de}, \label{eq:de_cont}
\end{align}
where $w_0$ is the dark energy equation of state parameter described as $w_0$= $p_{\rm de}/\rho_{\rm de}$, the dots represent derivative with respect to cosmic time $t$, and $H \equiv \dot{a}/a$ is the Hubble parameter. In this work, we consider two cases: in the $\xi $CDM model, $w_0$ is fixed at -1.001 (a phantom surrogate), while in the $\xi w$CDM model, it is treated as a free parameter. In the model name $\xi w$CDM, the symbol $w$ refers to this same parameter $w_0$.

\subsection{Background Evolution Equation}
Equation \eqref{eq:de_cont} is solved by rewriting it in terms of the scale factor $a$.  Using $\dot{\rho}=H\,d\rho/d\ln a$,
\begin{equation}
\frac{d\rho_{\rm de}}{d\ln a} + 3(1+w_0)\rho_{\rm de} = \xi \rho_{\rm de},
\end{equation}
hence
\begin{equation}
\frac{d\rho_{\rm de}}{d\ln a} = \bigl[\xi-3(1+w_0)\bigr]\rho_{\rm de}.
\end{equation}
Integrating from $a=1$ (today) to arbitrary $a$ gives
\begin{equation}
\rho_{\rm de}(a) = \rho_{\rm de,0}\; a^{\xi-3(1+w_0)}. \label{eq:rho_de_sol}
\end{equation}

For dark matter, Eq. \eqref{eq:dm_cont} becomes
\begin{equation}
\frac{d\rho_{\rm dm}}{d\ln a} + 3\rho_{\rm dm} = -\xi\rho_{\rm de}
= -\xi\rho_{\rm de,0}\; a^{\xi-3(1+w_0)}.
\end{equation}
Multiplying by the integrating factor $a^3$ yields
\begin{equation}
\frac{d}{d\ln a}\bigl(a^3\rho_{\rm dm}\bigr) = -\xi\rho_{\rm de,0}\; a^{\xi-3w_0}.
\end{equation}
Integrating from $a=1$:
\begin{equation}
a^3\rho_{\rm dm}(a) - \rho_{\rm dm,0} = -\xi\rho_{\rm de,0}\int_{1}^{a} a'^{\,\xi-3w_0-1}da'.\label{eq:deq}
\end{equation}
The integral evaluates to $\displaystyle \frac{a^{\xi-3w_0}-1}{\xi-3w_0}$ (provided $\xi\neq 3w_0$).  Thus
\begin{equation}
\rho_{\rm dm}(a) = \rho_{\rm dm,0}\,a^{-3} \;-\; \xi\rho_{\rm de,0}\,
\frac{a^{\xi-3w_0-3}-a^{-3}}{\xi-3w_0}. \label{eq:rho_dm_sol}
\end{equation}

\noindent When $\xi=0$, Eq.~\eqref{eq:rho_dm_sol} and Eq.~\eqref{eq:rho_de_sol} recover the uncoupled (standard) scaling laws $\rho_{\rm dm} \propto a^{-3}$ and $\rho_{\rm de} \propto a^{-3(1+w_0)}$, where physical energy densities decrease only due to volume expansion in cosmic time $t$. For $\xi < 0$, the interaction term $Q = -\xi H \rho_{\rm de} > 0$ shows continuous energy transfer from dark energy to dark matter~\cite{Amendola2000}. The ratio of the IDE dark-matter density, shown in equation \eqref{eq:rho_dm_sol}, to the standard \(\Lambda\)CDM expression ($\rho_{\rm dm,0}\,a^{-3}$) at the same scale factor ($a$) is \[\frac{\rho_{\rm dm}^{\rm IDE}(a)}{\rho_{\rm dm}^{\Lambda\rm CDM}(a)} = 1 + \frac{|\xi|\Omega_{\rm de,0}}{\Omega_{\rm dm,0}} \frac{a^{\xi-3w_0} - 1}{\xi - 3w_0}, \] which is less than unity for $a < 1$ on the stable branch \(\xi < 0\), \(w_0 < -1\). This confirms analytically that the comoving dark-matter density is lower in the past than in the standard \(\Lambda\)CDM model. At later times ($a \to \infty$), although energy flow continues to supply energy to the dark matter component, spatial expansion remains the dominant effect, guaranteeing  that the total physical matter density $\rho_{\rm dm}(a)$ and the fractional matter density parameter $\Omega_m(a)$ decrease during the accelerated expansion time.

\subsection{Modified Friedmann Equation}
\noindent The Friedmann equation for a flat universe containing baryons, radiation, and the interacting dark sector is given by \cite{Amendola2000, Majerotto2009}:
\begin{equation}
H^2(a) = H_0^2\Bigl[ \Omega_{b,0}a^{-3} + \Omega_{r,0}a^{-4}
+ \frac{\rho_{\rm dm}(a)}{\rho_{c,0}} + \frac{\rho_{\rm de}(a)}{\rho_{c,0}} \Bigr], \label{eq:friedmann}
\end{equation}
with $\rho_{c,0}=3H_0^2/(8\pi G)$ and $\Omega_{i,0}=\rho_{i,0}/\rho_{c,0}$.
Inserting equations \eqref{eq:rho_de_sol} and \eqref{eq:rho_dm_sol} to this equation, we obtain the explicit algebraic form
\begin{equation}
\begin{aligned}
H^2(a) = H_0^2\Bigl[ &\Omega_{b,0}a^{-3}+\Omega_{r,0}a^{-4} \\
&+ \Omega_{\rm dm,0}a^{-3}
- \xi\Omega_{\rm de,0}\, \frac{a^{\xi-3w_0-3}-a^{-3}}{\xi-3w_0} \\
&+ \Omega_{\rm de,0}\,a^{-3(1+w_0)+\xi} \Bigr].
\end{aligned}
 \label{eq:H2_algebraic}
\end{equation}
Equation \eqref{eq:H2_algebraic} is the core background evolution of our minimal interacting dark-energy model. It contains only \text{two additional parameters} beyond $\Lambda$CDM: the coupling strength $\xi$ and the dark energy equation of state $w_0$. The limit $\xi\to 0,\;w_0=-1$ reproduces the standard $\Lambda$CDM expansion history.

\subsection{Perturbation Equations}

\noindent We work in the synchronous gauge, following the formulation of 
Refs.~\cite{Ma1995,Gavela2010}. The perturbed metric is given by:

\begin{equation}
ds^2 = a^2(\tau)\left[-d\tau^2 + (\delta_{ij} + h_{ij})dx^i dx^j\right],
\end{equation}
where $h_{ij}$ is the metric perturbation. The energy-mementum transfer four vector is considered to be along the the dark matter four velocity,
\begin{equation}
Q_{dm}^{\mu} = -Q_{de}^{\mu} = -\xi H \rho_{de} u_{dm}^{\mu},
\end{equation}
so that there is no transfer of momentum in the rest frame of the dark matter thereby satisfying the equivalence principle  ~\cite{Majerotto2009,Gavela2010}. In this case, since the interaction rate is proportional to $H\rho_{\rm de}$, incorporating
\begin{equation}
Q_{\rm de}=\xi H\rho_{\rm de},
\end{equation}
it follows that the first-order perturbation includes the perturbation of the Hubble rate of expansion. 

\begin{equation}
\frac{\delta Q_{\rm de}}{Q_{\rm de}}
=
\delta_{\rm de}
+\frac{\delta H}{H}.
\end{equation}
The perturbation $\delta H$ in the synchronous gauge contributes to the metric and velocity perturbations in the perturbed conservation equations. It is essential to include this contribution in order to attain gauge invariant treatment of the interacting dark sector~\cite{Gavela2010}. In our work, we adopt $\xi<0$ and $w_0<-1$, consistent with the opposite sign convention of the interaction parameter employed in Ref.~\cite{Gavela2009}. Thus, the signs for the interaction terms in the perturbation equations are different from that stated in~\cite{Gavela2009}.  With this covariant formulation, the linear perturbation equations for the dark matter density contrast $\delta_{dm} \equiv \delta\rho_{dm}/\rho_{dm}$ 
and velocity divergence $\theta_{dm} \equiv ik^j v_{dm,j}$, and the 
corresponding dark energy perturbations $\delta_{de}$ and $\theta_{de}$, take the following form in the synchronous gauge~\cite{ Gavela2009, Gavela2010}:
\begin{align}
\delta_{\mathrm{dm}}' &= -\left(\theta_{\mathrm{dm}} + \frac{h'}{2}\right) 
- \xi \mathcal{H}\frac{\rho_{\mathrm{de}}}{\rho_{\mathrm{dm}}}\left(\delta_{\mathrm{de}} - \delta_{\mathrm{dm}}\right) 
- \xi \frac{\rho_{\mathrm{de}}}{\rho_{\mathrm{dm}}}\left(\frac{\theta_T}{3} + \frac{h'}{6}\right),
\label{eq:delta_dm}
\\
\theta_{\mathrm{dm}}' &= -\mathcal{H}\theta_{\mathrm{dm}},
\label{eq:theta_dm}
\\
\delta_{\mathrm{de}}' &= -(1 + w_0)\left(\theta_{\mathrm{de}} + \frac{h'}{2}\right) 
- 3\mathcal{H}(1 - w_0)\left[\delta_{\mathrm{de}} + \mathcal{H}\left(3(1 + w_0) - \xi\right)\frac{\theta_{\mathrm{de}}}{k^2}\right] 
+ \xi \left(\frac{\theta_T}{3} + \frac{h'}{6}\right),
\label{eq:delta_de} 
\\
\theta_{\mathrm{de}}' &= 2\mathcal{H}\left(1 - \frac{\xi}{1 + w_0}\right)\theta_{\mathrm{de}} 
+ \frac{k^2}{1 + w_0}\delta_{\mathrm{de}} 
+ \xi \mathcal{H}\frac{\theta_{\mathrm{dm}}}{1 + w_0},
\label{eq:theta_de}
\end{align}
where the primes denote derivatives with respect to conformal time $\tau$, 
$\mathcal{H} \equiv a'/a = a H$ is the conformal Hubble rate, $k$ is the comoving wavenumber, and $\theta_T$ is the total fluid velocity divergence.
The dark matter velocity perturbation equation remains unchanged from the 
standard uncoupled case due to our choice of energy-momentum transfer 
aligned with the dark matter four-velocity~\cite{Majerotto2009,Gavela2010}. \\
For our stable interacting system free from early-time non-adiabatic 
instabilities, we require a condition $\xi(1 + w_0) > 0$ which follows from the gauge-invariant analysis of the perturbation  equations~\cite{Gavela2009, Gavela2010, Jackson2009}. For a negative coupling ($\xi < 0$), this 
condition requires $w_0 < -1$, placing the model in the phantom regime. This choice guarantees that the perturbation equations are consistent at both early and late times, and prevents the growth of instabilities that would affect the interacting dark sector.

\section{Methodology and Data Analysis}
\label{sec:methods}
\noindent In this section, we will outline the statistical methodology and observational datasets utilized in our study, along with the implementation of the modified CLASS code used for the interacting dark energy models.
\subsection{Modified CLASS Implementation}
\noindent To carry out the interacting dark energy (IDE) model, we modified the CLASS Boltzmann code by implementing the dark sector dynamics characterized by the energy transfer rate Q = $\xi H \rho_{\rm de}$ \cite{Blas2011}. The adjustments are implemented in both the background and perturbation modules.\\
\noindent In the background module, we address the modified continuity equations for dark matter and dark energy densities, as presented in Eqs. \eqref{eq:dm_cont} and \eqref{eq:de_cont}. The evolution of energy density, described in Eqs.\eqref{eq:rho_de_sol} and \eqref{eq:rho_dm_sol}, is used to determine the Hubble parameter through the modified Friedmann equation (Eq. \eqref{eq:H2_algebraic}).\\
In the perturbation module, we implement the gauge-invariant evolution equations for the dark matter and dark energy density contrasts and velocity divergences in the synchronous gauge, following the formulation of Refs. ~\cite{Bruni1992, Gavela2010}. 
\subsection{Data and Likelihoods}
\noindent We employ a comprehensive combination of cosmological datasets chosen to provide complementary constraints on both background expansion and structure growth. 
\begin{itemize}
\item \textbf{Planck 2018 CMB:} We apply the full Planck 2018 likelihood that includes the high-$\ell$ TT, TE, and EE power spectra in the multipole range $30 \le \ell \le 2500$ and the low-$\ell$ (TT, EE) power spectra for $2 \le \ell \le 29$ as we well as the lensing construction power spectrum ~\cite{Planck2018VIII, Planck2020}. This is denoted as \text{Planck} throughout our dataset combinations.
\item \textbf{DESI DR2 BAO:} We incorporate the latest Baryon Acoustic Oscillation measurements from the Dark Energy Spectroscopic Instrument Data Release 2 (DESI DR2), 2024/2025 release ~\cite{DESI2024III, DESI2024IV, DESI2024VI, DESI2024EDR}. Seven effective redshift bins are covered by the isotropic and anisotropic BAO measurements in the DESI DR2 dataset: BGS at $z = 0.295$; LRG1 at $z = 0.510$; LRG2 at $z = 0.706$; a combined LRG3+ELG1 sample at $z = 0.934$; ELG2 at $z = 1.321$; QSO at $z = 1.484$; and Lyman-$\alpha$ forest tracers at $z = 2.330$. We employ the correlated triplet of isotropic ($D_V / r_{\mathrm{d}}$), transverse ($D_M / r_{\mathrm{d}}$), and radial ($D_H / r_{\mathrm{d}}$) BAO measurements for each tracer. This new BAO dataset breaks important degeneracies between the Hubble constant $H_0$, the matter density $\Omega_m$, and dark energy parameters by providing notable constraints on both the angular diameter distance $D_M(z)$ and the Hubble expansion rate $H(z)$.
\item \textbf{Pantheon+ SNe Ia:} For Type Ia supernovae, we employ the Pantheon+ compilation~\cite{Brout2022} consisting of 1701 light curves ($0.001 < z < 2.26$). 
\item \textbf{Weak Lensing (WL) and Inter-Parameter Dynamics:} We incorporate measurements of cosmic shear from the Kilo-Degree Survey (KiDS-1000) and the Dark Energy Survey Year 3 (DES-Y3)~\cite{Amon2022, Asgari2021}. Although WL do not measure $H_0$ directly, they provide precise, independent constraints on the total matter density ($\Omega_m$) and its clumpiness ($\sigma_8$) via $S_8 \equiv \sigma_8 (\Omega_m / 0.3)^{0.5}$, which makes them important in evaluating model performance against structure-growth tensions. These structure-growth measurements break degeneracies in the parameters of the dark sector in our joint likelihood analysis ($\text{CMB}+\text{BAO}+\text{SN}+\text{WL}$). Weak-lensing data provide an independent probe of matter clustering and thus geometric probes (such as BAO, SNIa, and CMB), strengthening the robustness of the inferred $H_0$ constraints and minimize possible dependence on any single dataset (preventing data selection bias).\\
\textit{Since the primary objective of this study is to examine the $H_0$ tension in the context of interacting dark energy models, a dedicated investigation of the $S_8$ tension is not pursued here, but is being carried out separately in an ongoing companion study. We emphasize that the WL data are used here only as an additional constraint on the dark-sector parameters.}

\end{itemize}
We perform our analysis using progressive data set combinations: \text{Planck}, \text{Planck + BAO}, \text{Planck + BAO + SN} and \text{Planck + BAO + SN + WL}. This systematic approach allows us to assess the effect of each probe on the parameter constraints. Hereafter, SN and BAO represent the Pantheon+ supernova and  DESI DR2 BAO data respectively.

\subsection{Applying Interacting Dark Energy Models}

\noindent We conduct the analysis considering two different IDE models ($\xi $CDM and $\xi w$CDM) based on the stability condition, $\xi(1+w_0)>0$, suggested in the literature~\cite{Majerotto2009,Jackson2009,Gavela2010}. Such a stability condition is necessary for the exclusion of the non-adiabatic instabilities in the early universe.\\
In the first scenario, we analyze the $\xi w$CDM model, where both the interaction strength $\xi$ and the dark energy equation of state $w_0$ are simultaneously free parameters, satisfying the stability condition 
$\xi(1+w_0) > 0$~\cite{Jackson2009}. We use flat priors on $\xi$ between $[-0.85, 0.0]$ and $w_0$ between $[-1.2, -1.0]$. The negative coupling implies that energy flows from dark energy to dark matter ($Q = \xi H \rho_{\rm de} < 0$), while the phantom equation of state $w_0 < -1$ satisfies the stability condition $\xi(1+w_0)>0$ ~\cite{Jackson2009}. This setup permits the dark sector interaction to explore the entire phantom regime, as required by the theoretical stability conditions.\\
In the second scenario, we consider the $\xi$CDM model, which is a stable surrogate for a coupled vacuum model. We fix $w_{0} = -1.001$ in the phantom regime and allow the coupling $\xi$ to vary freely within the prior range $[-0.85,0.0]$. The exact vacuum models (where $w_0=-1$) with $\rho_{de}$-proportional coupling lead to significant instabilities in dark energy pressure perturbations ($\delta p_{de} \propto \delta\rho_{de}/(1+w_0)$). Fixing $w_{0} = -1.001$, we regularize the stability of CLASS numerical implementations by maintaining an approximate-vacuum background evolution and satisfying $\xi(1+w_0)>0$. We apply this technique to distinguish the phenomenological effects of the dark matter-dark energy interaction strength compared to the impact of a dynamical dark energy equation of state. Therefore the $\xi$CDM model has just one additional parameter ($\xi$) compared to $\Lambda$CDM, making it a more conservative test of the interacting dark sector hypothesis.
\subsection{Parameter Estimation and Priors}
\noindent We describe the IDE model with six cosmological parameters: $H_0$, $\Omega_b h^2$, $\Omega_c h^2$, $\ln(10^{10}A_s)$, $n_s$, and $\tau$, together with the two newly added external parameters ($\xi$, $w_0$), making the total parameter eight. The $\Omega_b h^2$ and $\Omega_c h^2$ describe the physical density of baryons and cold dark matter, $H_0$ is the Hubble constant today, $A_s$ and $n_s$ are the primordial amplitude and spectral index, and $\tau$ is the optical depth to reionization. The two new free parameters, strength of the 
\begin{table}[H]
\centering
\small
\caption{Flat prior ranges for the cosmological parameters.}
\begin{tabular}{l c c}
    \hline
    Parameter & Prior Range & Model \\
    \hline
    $H_0$ [km s$^{-1}$ Mpc$^{-1}$] & $[50, 90]$ & $\xi$CDM and $\xi w$CDM \\
    $\Omega_b h^2$ & $[0.015, 0.030]$ & $\xi$CDM and $\xi w$CDM  \\
    $\Omega_c h^2$ & $[0.05, 0.25]$ & $\xi$CDM and $\xi w$CDM  \\
    $\ln(10^{10}A_s)$ & $[2.5, 3.5]$ & $\xi$CDM and $\xi w$CDM  \\
    $n_s$ & $[0.80, 1.1]$ & $\xi$CDM and $\xi w$CDM  \\
    $\tau_{\rm reio}$ & $[0.01, 0.15]$ & $\xi$CDM and $\xi w$CDM  \\
    $\xi$ & $[-0.85, 0.0)$ & $\xi$CDM and $\xi w$CDM  \\
    $w_0$ & $-1.001$ (fixed) & $\xi$CDM only\\
    $w_0$ & $[-1.2, -1.0]$ & $\xi w$CDM only\\
    \hline
    \hline
\end{tabular}
\label{tab:priors}
\end{table}

\footnotetext[1] {The stability requirement $\xi(1+w_0) > 0$ is imposed on both models ($\xi$CDM and $\xi w$CDM). For $\xi < 0$ and $w_0 < -1$ (which holds $w_0 = -1.001$ for $\xi$CDM and $w_0 \in [-1.2, -1.0)$ for $\xi w$CDM), this condition is automatically satisfied.}

\noindent interaction $\xi$ and the present-day dark energy equation of state $w_0$ characterize the IDE model. Sampling of $H_0$ is done directly, making $H_0-\xi$ degeneracy clearly visible and consistent with a few investigations of interacting dark energy analysis~\cite{DiValentino2020b, Salvatelli2013}.
\subsubsection{MCMC Implementation}
\noindent To generate samples from the posterior distributions, we use the \textsc{Cobaya} framework~\cite{Torrado2021}, a Bayesian analysis code that interfaces with the \textsc{CLASS} Boltzmann solver via the \textsc{Classy} wrapper. The \textsc{Cobaya} sampler employs the affine-invariant MCMC ensemble sampler \textsc{emcee}~\cite{Foreman-Mackey2013} for efficient exploration of the parameter space. Convergence was assessed using the Gelman-Rubin diagnostic, requiring   $\hat{R}-1 < 0.02$ for all sampled parameters~\cite{Gelman1992}. The median with the 16th and 84th percentiles (68\% credible interval) are provided as the final parameter constraints. 

\subsubsection{Model Comparison}
\noindent The IDE parameter space is constrained via Bayesian inference. Given data $\mathcal{D}$, the posterior probability distribution $P(\theta | \mathcal{D})$ for parameter vector $\theta$ adheres to Bayes' theorem:
\begin{equation}
	P(\theta | \mathcal{D}) = \frac{\mathcal{L}(\mathcal{D}|\theta)\pi(\theta)}{\mathcal{Z}(\mathcal{D})},
\end{equation}
where $\pi(\theta)$ is the prior probability, $\mathcal{L}(\mathcal{D}|\theta)$ is the likelihood function, and $\mathcal{Z}(\mathcal{D})$ is the Bayesian evidence.\\
The priors for every free parameter in the IDE analysis are compiled in Table~\ref{tab:priors}. Over the selected ranges, all parameters maintain flat priors. \\
We used the the Akaike Information Criterion (AIC)~\cite{Akaike1974} and Bayesian Information Criterion (BIC)~\cite{Schwarz1978} to compare models:
\begin{align}
\mathrm{AIC} &= -2\ln\mathcal{L}_{\mathrm{max}} + 2k, \label{eq:aic}\\
\mathrm{BIC} &= -2\ln\mathcal{L}_{\mathrm{max}} + k\ln N, \label{eq:bic}
\end{align}
where $k$ stands for the number of free parameters, $N$ is the sum of the effective numbers of independent data points across the combination, and $\mathcal{L}_{\mathrm{max}}$ is the maximum likelihood. We define $\Delta\mathrm{AIC} \equiv \mathrm{AIC}_{IDE} - \mathrm{AIC}_{\Lambda\mathrm{CDM}}$. The negative values here favor the IDE model.\\
The strength of statistical evidence favoring one model over another is interpreted using standard empirical scales~\cite{Burnham2002, Kass1995}

\begin{itemize}
	\item \textbf{AIC}: $|\Delta| < 2$ (indistinguishable), $2$--$4$ (positive), $4$--$10$ (strong), $\geq 10$ (very strong)~\cite{Burnham2002};
	\item \textbf{BIC}: $|\Delta| < 2$ (weak), $2$--$6$ (positive), $6$--$10$ (strong), $\geq 10$ (Decisive / Overwhelming evidence).
\end{itemize}
We emphasize that all statements about model preference in this work should be interpreted within the context of these conventions. 
\subsubsection{Hubble Tension Quantification}
We employ the Hubble tension using the Gaussian tension metric in accordance with~\cite{Verde2019}:
{\footnotesize
	\begin{align}
		T_{\mathrm{SH0ES\text{-}Planck}} &= \frac{|H_0^{\mathrm{SH0ES}} - H_0^{\mathrm{Planck}}|}{\sqrt{\sigma_{\mathrm{SH0ES}}^2 + \sigma_{\mathrm{Planck}}^2}}, \label{eq:tension_planck}\\[-3pt]   
        \end{align}
        \begin{align}
		T_{IDE\text{-}SH0ES} &= \frac{|H_0^{IDE} - H_0^{\mathrm{SH0ES}}|}{\sqrt{\sigma_{IDE}^2 + \sigma_{\mathrm{SH0ES}}^2}}. \label{eq:tension_model}
	\end{align}
}

The reduction in fractional tension attained by IDE is then given as:
{\footnotesize
	\begin{equation}
		\mathcal{R} = 1 - \frac{T_{IDE{\text{-}SH0ES}}}{T_{\mathrm{SH0ES\text{-}Planck}}}. \label{eq:reduction}
	\end{equation}
}
\vspace{-0.75cm}
\section{Results}
\vspace{-0.3cm}
\label{sec:results}
\noindent In this section, we present the constraints and main results for the interacting dark energy models introduced in Sec.\ref{sec:methods} using four systematic observational dataset combinations: Planck, Planck +  BAO(DESI DR2), Planck + BAO(DESI DR2) + SN(Pantheon+), and Planck + BAO(DESI DR2) + SN(Pantheon+) + WL. This progressive approach enables us to evaluate how each observational probe modifies the constraints on the model parameters and to assess the reliability of any possible evidence for dark sector interactions. Unless otherwise specified, all constraints are reported at the 68\% confidence level. 
\subsection{Constraints on $\xi$CDM Model and the Constraining Power of DESI DR2 BAO Data}
\label{sec:constraining power}

\noindent The constraints for the $\xi$CDM model at the 68\% confidence level (C.L.) are presented in Table~\ref{tab:constraints_xi}, where the equation of state for dark energy is specified to $w_0 = -1.001$ (phantom-like surrogate), while the coupling parameter $\xi$ is allowed to be free.\\
\noindent From the table, we observe that the Hubble constant in $\xi$CDM model is higher compared to the standard $\Lambda$CDM model. For the Planck-only dataset, we obtain $H_0 = 71.59^{+1.69}_{-2.30}$ km s$^{-1}$ Mpc$^{-1}$, which is substantially higher than the standard $\Lambda$CDM value of $67.4 \pm 0.5$ km s$^{-1}$ Mpc$^{-1}$. This improvement arises from the effect of the negative interaction strength. Based on our sign convention (see Eq.~\eqref{eq:dm_cont} and Eq.~\eqref{eq:de_cont}), the negative coupling represents a continuous energy transfer from dark energy to dark matter. This energy transfer leads to a deviation of the dark-matter density from the standard $a^{-3}$ dilution law, which modifies the evolution of dark matter from the early to the late universe.

\begin{table}[H]
\centering
\small
\caption{Constraints (68\% C.L.) on cosmological parameters for the interacting dark energy model ($\xi $CDM) using different observational data sets. The coupling parameter $\xi$ is free while $w_0 = -1.001$ is kept constant. The statistics for model comparison are determined with respect to the associated $\Lambda$CDM model, such that $\Delta X = X_{\mathrm{IDE}} - X_{\Lambda\mathrm{CDM}}$.}

\label{tab:constraints_xi}
\vspace{0.2cm}
\renewcommand{\arraystretch}{1.3}

\begin{tabular}{@{} l c c c c @{}}
\hline
\hline
\textbf{Parameter} & 
\textbf{\scriptsize Planck} &          
\textbf{\scriptsize Planck + BAO} &  
\textbf{\scriptsize Planck + BAO + SN} &  
\textbf{\scriptsize Planck+BAO+SN+WL}   \\
\hline










$\omega_b = \Omega_b h^2$ &
$0.0224^{+0.0001}_{-0.0002}$ &
$0.0224^{+0.0002}_{-0.0002}$ &
$0.0224^{+0.0001}_{-0.0001}$ &
$0.0224^{+0.0001}_{-0.0001}$ \\

$\omega_c = \Omega_c h^2$ &
$0.1199^{+0.0014}_{-0.0012}$ &
$0.1195^{+0.0010}_{-0.0010}$ &
$0.1196^{+0.0009}_{-0.0009}$ &
$0.1195^{+0.0009}_{-0.0009}$ \\

$H_0$ (km s$^{-1}$ Mpc$^{-1}$) &
$71.5941^{+1.6930}_{-2.3004}$ &
$70.1813^{+0.5426}_{-0.4315}$ &
$69.7630^{+0.5646}_{-0.4398}$ &
$69.7217^{+0.5018}_{-0.4074}$ \\

$\ln(10^{10} A_s)$ &
$3.0375^{+0.0124}_{-0.0086}$ &
$3.0442^{+0.0138}_{-0.0140}$ &
$3.0449^{+0.0114}_{-0.0112}$ &
$3.0449^{+0.0120}_{-0.0111}$ \\

$n_s$ &
$0.9661^{+0.0039}_{-0.0064}$ &
$0.9657^{+0.0041}_{-0.0037}$ &
$0.9658^{+0.0030}_{-0.0030}$ &
$0.9659^{+0.0034}_{-0.0033}$ \\

$\tau_{\mathrm{reio}}$ &
$0.0532^{+0.0039}_{-0.0041}$ &
$0.0547^{+0.0068}_{-0.0064}$ &
$0.0555^{+0.0061}_{-0.0057}$ &
$0.0555^{+0.0058}_{-0.0059}$ \\

$\xi$ &
$-0.2962^{+0.1319}_{-0.1927}$ &
$-0.1152^{+0.0178}_{-0.0301}$ &
$-0.0760^{+0.0185}_{-0.0303}$ &
$-0.0740^{+0.0169}_{-0.0291}$ \\

$\Omega_m$ &
$0.2780^{+0.0194}_{-0.0125}$ &
$0.2882^{+0.0038}_{-0.0054}$ &
$0.2916^{+0.0042}_{-0.0050}$ &
$0.2920^{+0.0037}_{-0.0046}$ \\

$\Omega_\Lambda$ &
$0.7220^{+0.0125}_{-0.0194}$ &
$0.7118^{+0.0054}_{-0.0038}$ &
$0.7084^{+0.0050}_{-0.0042}$ &
$0.7080^{+0.0046}_{-0.0037}$ \\

$\sigma_8$ &
$0.8687^{+0.0223}_{-0.0316}$ &
$0.8512^{+0.0088}_{-0.0060}$ &
$0.8458^{+0.0080}_{-0.0065}$ &
$0.8452^{+0.0074}_{-0.0057}$ \\
\hline
$\Delta\chi^2$ &
 &
$-0.83$ &
$-5.80$ &
$-7.00$ \\

$\Delta\mathrm{AIC}$ &
 &
$+1.17$ &
$-3.80$ &
$-5.00$ \\

$\Delta\mathrm{BIC}$ &
 &
$+2.11$ &
$+1.65$ &
$+0.45$ \\
\bottomrule
\end{tabular}
\end{table}

\begin{figure}[H]
	\centering
	\includegraphics[width=0.75\textwidth]{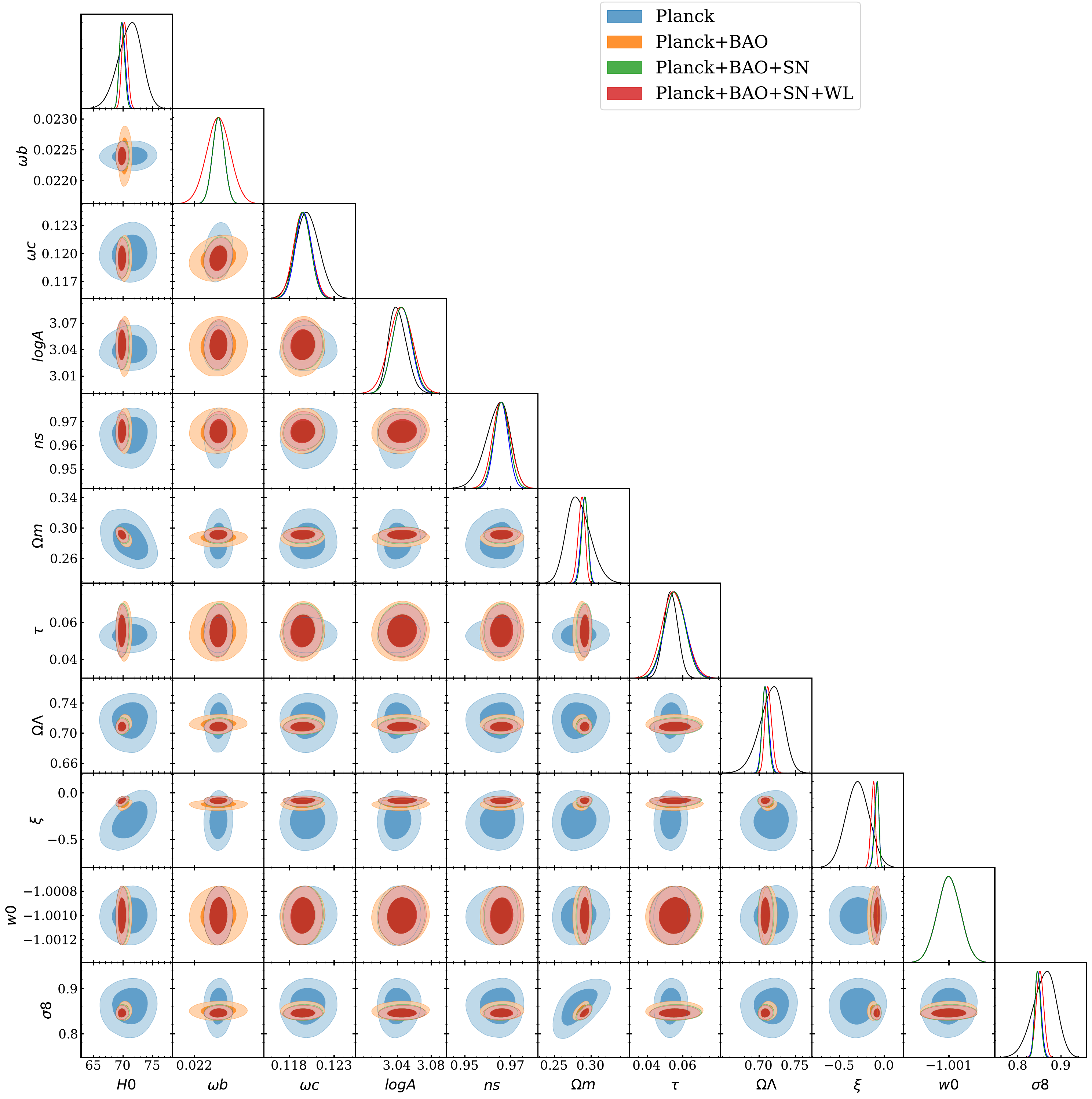}
	\caption{Triangle (corner) plot for the $\xi $CDM model illustrating the one dimensional marginalized posterior distributions (diagonal) and two dimensional marginalized posterior distributions (off diagonal) at 68\% and 95\% confidence level (C.L) for the cosmological parameters obtained from the four datasets: Planck (blue), Planck+BAO (orange), Planck+BAO+SN (green), and Planck+BAO+SN+WL (red). The contours display the 68\% (inner) and 95\% (outer) confidence regions, with the best fit values indicated by the peak of each distribution. The inclusion of BAO, SN, and WL data progressively tightens the parameter constraints, particularly for $H_0$, $\Omega_m$, and $\xi$, while shifting the best fit values toward lower $H_0$ and $\Omega_m$ and $\xi$.}
\label{fig:corner_plot}
\end{figure}

\begin{figure}[H]
	\centering
	\includegraphics[width=1.0\textwidth]{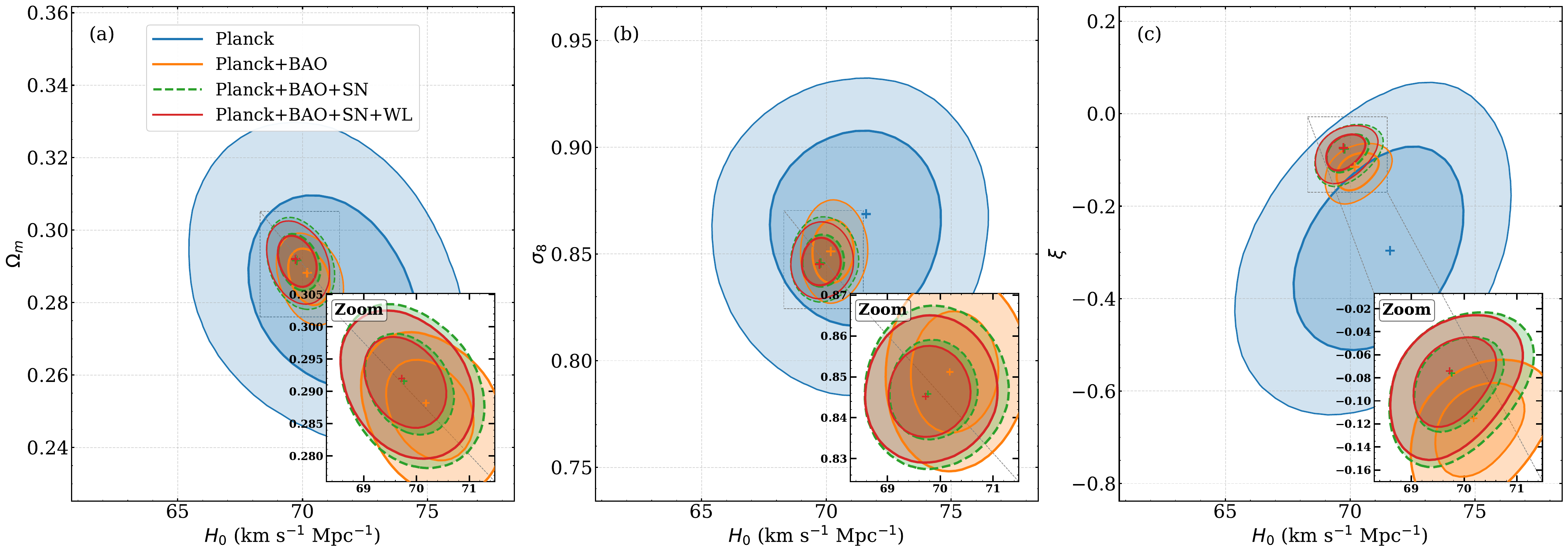}
	\caption{Two dimensional marginalized posterior probability distributions for the parameter pairs (a) $(H_0, \Omega_m)$, (b) $(H_0, \sigma_8)$, and (c) $(H_0, \xi)$ obtained from the four data combinations conducted in this work: planck-only (blue), Planck+BAO (orange), Planck+BAO+SN (green), and Planck+BAO+SN+WL (red). The filled contours represent the 68\% (inner, darker) and 95\% (outer, lighter) confidence regions, with corresponding contour lines overlaid. For each dataset, the best fit values are marked by '+' sign. The zoom insets in each panel focus on the combined datasets (excluding Planck-only) to clearly show the progressively tighter constraints achieved with the inclusion of BAO, SN, and WL data.}
\label{fig:contours}
\end{figure}

\noindent Specifically, although the dark-matter density continues to decrease due to the expansion of the universe, its evolution is different from the standard $a^{-3}$ scaling because of the continuous energy flow from dark energy to dark matter, as stated by Eq.~\eqref{eq:rho_dm_sol}. Hence, to keep the current value of physical cold dark matter density $\Omega_c h^2$, fixed from the CMB acoustic peaks, the model shifts the inferred present-day Hubble constant $H_0$ toward higher value thereby alleviating the $H_0$ tension. This behavior  is indeed a clear signature of the degeneracy between $\xi$ and $H_0$ visible in Figure~\ref{fig:corner_plot} and more prominently  in Figure~\ref{fig:contours} (c), where more negative values of $\xi$ are associated with higher $H_0$ values. The resulting shift in the $H_0$ posterior distributions with respect to  SH0ES across all dataset combinations is demonstrated in figure~\ref{fig:H0_comparison}(a) for the $\xi\Lambda\text{CDM}$ model. Notably, $\xi\Lambda\text{CDM}$ achieves a significant tension reduction in the $H_0$ tension, to $\approx 2.8\sigma$, supported by a strongly favorable AIC($\Delta\text{AIC} = -5.00$). While $\Delta\text{BIC} = +0.45$ gives a weak preference to $\Lambda\text{CDM}$ model due to the Bayesian penalty for the additional parameter $\xi$, the value $\vert{}\Delta\text{BIC}\vert{} < 2$ shows that both models fit the data with comparable statistical support.\\

\noindent In agreement with the stability condition $\xi(1+w_0)>0$, the coupling $\xi$ must be negative for all dataset combinations. We get  $\xi = -0.296^{+0.132}_{-0.193}$ for the Planck-only dataset, which prefers a non zero coupling at about $2.2\sigma$. But the inclusion of BAO data tightens the constraint significantly, providing $\xi = -0.115^{+0.018}_{-0.030}$ and reducing the central value to a very small magnitude. Adding SN and weak lensing data further shifts the coupling to $\xi \approx -0.074$, showing that the interaction remains with non-zero negative values but becomes smaller in size. This residual negative coupling neverthless continues to contribute to the high $H_0$ value, indicating that even a small interaction can have observable impact on the expansion history. This behavior is clearly evident in Figure~\ref{fig:contours} (c), where the inclusion of BAO data greatly shrinks the allowed parameter region.\\
\noindent In such way, the full combination of datasets (Planck + BAO + SN + WL) gives $H_0 = 69.72^{+0.50}_{-0.41}$ km s$^{-1}$ Mpc$^{-1}$. We quantify the tension between the IDE-inferred $H_0$ and the SH0ES measurement of $H_0 = 73.04 \pm 1.04$ km s$^{-1}$ Mpc$^{-1}$~\cite{Riess2022} using Eqs.~\eqref{eq:tension_planck} and \eqref{eq:tension_model}. The standard $\Lambda$CDM early-universe inference gives $T_{\mathrm{SH0ES-Planck}} \sim 5.0\sigma$. In the $\xi$CDM model, the descrepancy between the IDE-inferred $H_0$ and SH0ES is reduced to $T_{\mathrm{IDE-SH0ES}} \approx 2.8\sigma$, which corresponds to a fractional reduction of $\mathcal{R} \approx 44\%$ (by Eq.~\eqref{eq:reduction}). This confirms that the $\xi$CDM model can alleviate the $H_0$ tension, although it does not resolve it entirely.\\
The constraining power of the DESI DR2 BAO data is particularly effective in breaking the parameter degeneracies associated with interacting dark energy models. Figure~\ref{fig:corner_plot} visualizes this exceptional tightening of constraints in the $\xi$CDM model. In this model, the constraints are markedly tightened by the inclusion of DESI DR2 BAO measurements compared to Planck-only (blue contours), especially in the $(\xi, H_0)$, $(w_0, H_0)$, $(H_0, \Omega_m)$, and $(H_0, \Omega_\Lambda)$ planes. The most remarkable feature is the pronounced tightening of the $(\xi, H_0)$  degeneracy in the presence of DESI DR2 BAO data. 

\noindent The shift in the inferred $H_0$ value when incorporating BAO and SN data does not originate from a direct constraint on $H_0$ itself, but rather arises because these probes tightly constrain the late-time expansion history. While Planck CMB mainly constrain the physical dark matter density $\Omega_c h^2$, leaving a broad  degeneracy between $H_0$ and $\xi$ that allows higher $H_0$ value, BAO and SN measure the expansion rate $H_0$ and distance scales $D_A(z)/D_L(z)$ across low to intermediate redshifts ($0 \lesssim z \lesssim 2.5$). This geometric anchor constrains the parameter space for $\xi$ at late times, thereby reducing the $H_0$ value compared to the Planck-only posterior.\\
In the $\xi$CDM model, the presence of a negative coupling parameter causes the modification of the background expansion history $H(z)$, affecting the redshift evolution of dark matter density, which in turn alters the shape of $H(z)$ in comparison with $\Lambda$CDM. The high value of $H(z)$ from the Planck-alone degeneracy predict that there is a rapid late-time expansion rate $H(z)$ which is not consistent with the actual distance measurements obtained from BAO and Supernovae. When these data are included, the allowed region in the ($H_0$, $\xi$) plane contracts as only the low value of $H_0$ can simultaneously fit both CMB and geometric measurements. Therefore, the decrease in $H_0$ is a geometric consistency requirement: the data force the model to choose an expansion history that is consistent with both early-and late-time measurements. \\
\noindent As shown from Table~\ref{tab:constraints_xi},  $H_0$ attains a value of $71.59^{+1.69}_{-2.30}$ km s$^{-1}$ Mpc$^{-1}$ in the Planck-only case, with the 68\% C.L. posterior covering from $69.3$ to $73.3$ km s$^{-1}$ Mpc$^{-1}$, and tightens substantially to $H_0 = 70.18^{+0.54}_{-0.43}$ km s$^{-1}$ Mpc$^{-1}$ for the Planck + BAO combination, and further to $H_0 = 69.72^{+0.50}_{-0.41}$ km s$^{-1}$ Mpc$^{-1}$ for the full dataset combination. This results in a reduction of the $H_0$ uncertainty from approximately $4.0~\mathrm{km\,s^{-1}\,Mpc^{-1}}$ (Planck-only) to about $0.9~\mathrm{km\,s^{-1}\,Mpc^{-1}}$ (Planck + BAO + SN + WL), an improvement by a factor of more than four. This degeneracy occurs because a negative coupling $\xi$ reduces the dark matter density, thereby necessitating a larger $H_0$ to preserve the $\Omega_c h^2$ value fixed by the CMB acoustic peaks. The inclusion of DESI DR2 BAO data efficiently breaks this degeneracy by constraining the late-time evolution of the universe.\\
In the $\xi$CDM model, the Planck-only contours also exhibit a prominent anti-correlation between $H_0$ and $\Omega_m$, representing a well-known geometrical degeneracy. This degeneracy is also significantly diminished when DESI DR2 BAO data are included. The value of $\Omega_m$ constraint improves from $0.278^{+0.019}_{-0.013}$ (Planck-only) to $0.292^{+0.004}_{-0.005}$ (Planck + BAO + SN + WL), an improvement of more than threefold. Similarly, $\Omega_\Lambda$ improves from $0.722^{+0.013}_{-0.019}$ to $0.708^{+0.005}_{-0.005}$, an improvement more than a factor of three.\\
The model comparison statistics gives important assessment to see whether the additional parameter $\xi$ is justified by the data. For the Planck + BAO dataset combination, we get $\Delta\chi^2 = -0.83$, which is very small. The $\Delta\mathrm{AIC} = +1.17$ and $\Delta\mathrm{BIC} = +2.11$ are both positive, revealing that the $\Lambda$CDM model is preferred over the $\xi$CDM model when the extra parameter is penalized. Based on the AIC scale~\cite{Burnham2002}, a value of $|\Delta\mathrm{AIC}| < 2$ shows that the $\Lambda$CDM  and $\xi$CDM models are indistinguishable. On the other hand, we get $\Delta\chi^2 = -5.80$, for the Planck + BAO + SN dataset combination, indicating a moderate improvement in fit. This gives  $\Delta\mathrm{AIC} = -3.80$, suggesting  a \emph{positive} preference for the $\xi$CDM model according to the AIC scale. But the $\Delta\mathrm{BIC} = +1.65$ yields only weak evidence in favor of this model. Based on the BIC perspective, values of $|\Delta\mathrm{BIC}| < 2$ demonstrate weak evidence, indicating that the $\Lambda$CDM and $\xi$CDM models are not clearly distinguishable for this dataset combination.\\
The full dataset integration (\text{Planck + BAO + SN + WL}), provides $\Delta\chi^2 = -7.00$, $\Delta\mathrm{AIC} = -5.00$, and $\Delta\mathrm{BIC} = +0.45$. Here, although the positive $\Delta\mathrm{BIC}$ shows a weak preference for $\Lambda\mathrm{CDM}$ due to the parameter penalty, the strong support from the AIC criterion ($\Delta\mathrm{AIC} = -5.00$) alongside the substantial improvement in fit ($\Delta\chi^2 = -7.00$) together provide compelling evidence in favor of the $\xi\mathrm{CDM}$ model.

\subsection{Constraints on the $\xi w$CDM Model}
\vspace{-0.05cm}

\begin{figure}[H]
	\centering
	\includegraphics[width=0.75\textwidth]{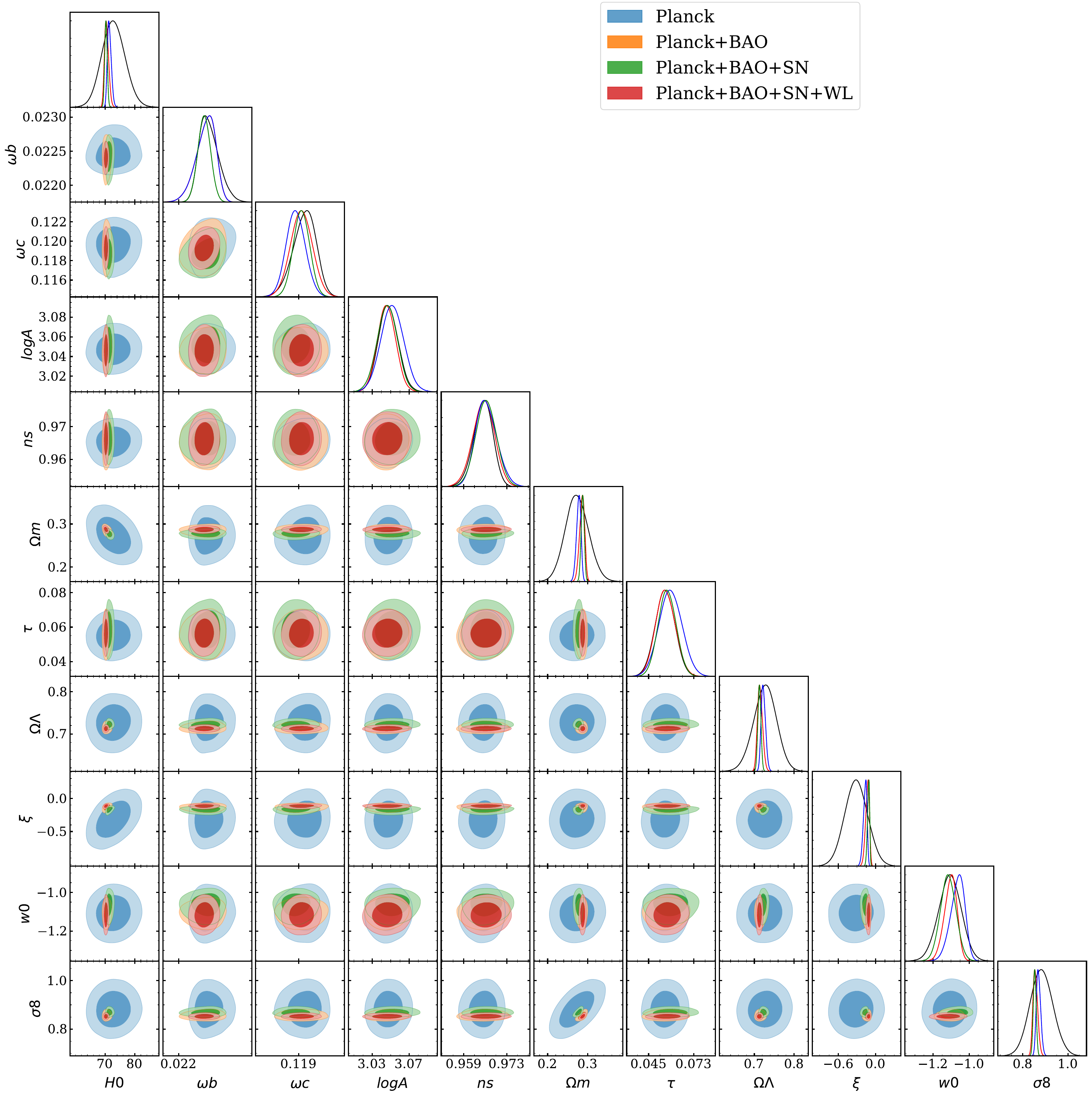}
	\caption{Triangle plot for the $\xi w$CDM model showing the one-dimensional marginalized posterior distributions (diagonal) and two-dimensional 68\% and 95\% confidence contours (off-diagonal) for all cosmological parameters, for Planck (blue), Planck+BAO (orange), Planck+BAO+SN (green), and Planck+BAO+SN+WL (red). The parameters shown are $H_0$, $\omega_b$, $\omega_c$, $\ln(10^{10}A_s)$, $n_s$, $\Omega_m$, $\tau_{\rm reio}$, $\Omega_\Lambda$, $\xi$, $w_0$, and $\sigma_8$. The progressive addition of external datasets (BAO, SN, and WL) visibly tightens the constraints and reduces degeneracies, particularly in the $H_0$-$\Omega_m$, $H_0$-$\xi$, and $\Omega_m$-$\sigma_8$ planes, highlighting the complementary nature of these probes.}
\label{fig:corner_plot2}
\end{figure}

\noindent A direct physical motivation behind the shift of the Hubble constant in the $\xi w$CDM model can be provided by the background evolution equations. In this case, the dark energy equation of state $w_0$ plays a role similar to the coupling $\xi$; where a highly negative (phantom) value of $w_0$ causes the dark energy density to decay more slowly with expansion, leading to a smaller value of matter density today for a given $\Omega_c h^2$ determined from CMB. To compensate, the model necessitates a higher $H_0$ to keep the product $\Omega_c h^2$ unchanged. This results in a strong degeneracy between $w_0$ and $H_0$, as shown in figure~\ref{fig:corner_plot2} in the $(w_0, H_0)$ space.

\noindent Table~\ref{tab:constraints_xi_w0} displays the 68\% confidence level constraints on the $\xi w$CDM model with both the coupling $\xi$ and dark energy equation of state $w_0$  are permitted to vary freely. 

\begin{table}[H]
\centering
\small
\caption{Constraints (68\% C.L.) on cosmological parameters of the interacting dark energy model ($\xi w$CDM model) from different observational data combinations. Both $\xi$ and $w_0$ are free. Model comparison statistics are determined with respect to the corresponding $\Lambda$CDM model, where $\Delta X = X_{\mathrm{IDE}} - X_{\Lambda\mathrm{CDM}}$. }
\label{tab:constraints_xi_w0}
\vspace{0.3cm}
\renewcommand{\arraystretch}{1.3}

\begin{tabular}{@{} l c c c c @{}}
\hline
\hline
\textbf{Parameter} & 
\textbf{\scriptsize Planck} &          
\textbf{\scriptsize Planck + BAO} &  
\textbf{\scriptsize Planck + BAO + SN} &  
\textbf{\scriptsize Planck+BAO+SN+WL}   \\
\hline

$\omega_b = \Omega_b h^2$ &
$0.0224^{+0.0002}_{-0.0001}$ &
$0.0225^{+0.0001}_{-0.0002}$ &
$0.0225^{+0.0001}_{-0.0002}$ &
$0.0224^{+0.0001}_{-0.0001}$ \\

$\omega_c = \Omega_c h^2$ &
$0.1200^{+0.0010}_{-0.0015}$ &
$0.1193^{+0.0012}_{-0.0012}$ &
$0.1186^{+0.0011}_{-0.0010}$ &
$0.1193^{+0.0009}_{-0.0009}$ \\

$H_0$ (km s$^{-1}$ Mpc$^{-1}$) &
$72.4550^{+4.0464}_{-3.6609}$ &
$70.2045^{+0.8684}_{-0.5164}$ &
$71.1034^{+0.8271}_{-0.4456}$ &
$70.2166^{+0.4451}_{-0.4003}$ \\

$\ln(10^{10} A_s)$ &
$3.0463^{+0.0113}_{-0.0099}$ &
$3.0451^{+0.0102}_{-0.0095}$ &
$3.0512^{+0.0127}_{-0.0120}$ &
$3.0465^{+0.0107}_{-0.0110}$ \\

$n_s$ &
$0.9659^{+0.0027}_{-0.0035}$ &
$0.9658^{+0.0033}_{-0.0037}$ &
$0.9655^{+0.0040}_{-0.0029}$ &
$0.9662^{+0.0034}_{-0.0033}$ \\

$\tau_{\mathrm{reio}}$ &
$0.0553^{+0.0060}_{-0.0061}$ &
$0.0551^{+0.0063}_{-0.0058}$ &
$0.0584^{+0.0072}_{-0.0070}$ &
$0.0564^{+0.0056}_{-0.0055}$ \\

$\xi$ &
$-0.3165^{+0.1875}_{-0.1824}$ &
$-0.1169^{+0.0206}_{-0.0405}$ &
$-0.1533^{+0.0185}_{-0.0402}$ &
$-0.1066^{+0.0121}_{-0.0228}$ \\

$w_0$ &
$-1.1031^{+0.0600}_{-0.0646}$ &
$-1.0927^{+0.0274}_{-0.0413}$ &
$-1.0488^{+0.0286}_{-0.0490}$ &
$-1.1164^{+0.0433}_{-0.0427}$ \\

$\Omega_m$ &
$0.2708^{+0.0301}_{-0.0271}$ &
$0.2875^{+0.0051}_{-0.0075}$ &
$0.2790^{+0.0043}_{-0.0066}$ &
$0.2874^{+0.0038}_{-0.0040}$ \\

$\Omega_\Lambda$ &
$0.7292^{+0.0271}_{-0.0301}$ &
$0.7125^{+0.0075}_{-0.0051}$ &
$0.7210^{+0.0066}_{-0.0043}$ &
$0.7126^{+0.0040}_{-0.0038}$ \\

$\sigma_8$ &
$0.8802^{+0.0517}_{-0.0487}$ &
$0.8521^{+0.0123}_{-0.0081}$ &
$0.8662^{+0.0114}_{-0.0072}$ &
$0.8523^{+0.0065}_{-0.0061}$ \\

\hline
\midrule
$\Delta\chi^2$ &
 &
$-0.69$ &
$-14.60$ &
$-6.67$ \\

$\Delta\mathrm{AIC}$ &
 &
$+3.31$ &
$-10.60$ &
$-2.67$ \\

$\Delta\mathrm{BIC}$ &
 &
$+5.20$ &
$+0.30$ &
$+8.23$ \\
\bottomrule
\end{tabular}
\end{table}
\noindent This phenomenon, combined with a negative coupling ($\xi < 0$), leads to a dual process resulting in higher $H_0$ values. Specifically, the alleviation of the $H_0$ tension within the $\xi w \text{CDM}$ frame work is based on the physically coupled mechanism governed by the negative value of the coupling parameter ($\xi < 0$) along with the phantom equation of state ($w_0 < -1$). 
\begin{figure}[H]
        \centering
        \includegraphics[width=0.95\textwidth]{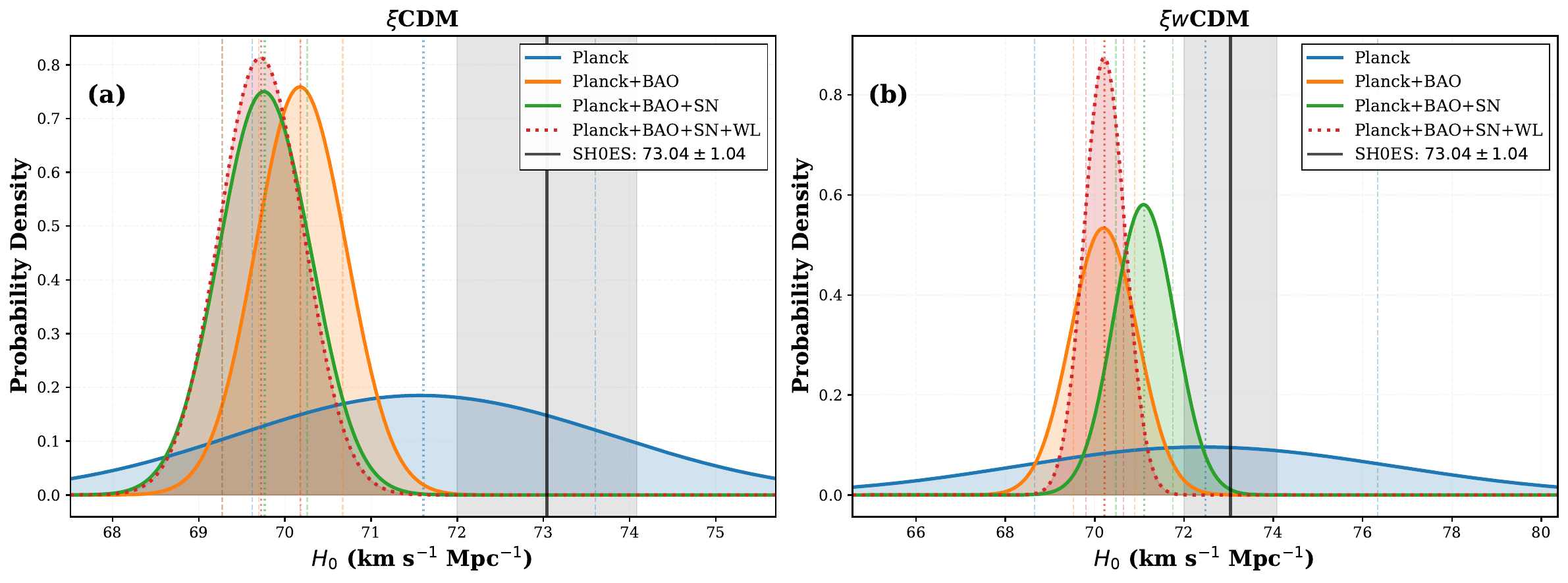}
        \caption{One-dimensional marginalized posterior probability distributions for the Hubble constant $H_0$ obtained from the four dataset combinations: Planck (blue), Planck+BAO (orange), Planck+BAO+SN (green), and Planck+BAO+SN+WL (red). Panel (a) shows the $\xi\Lambda\text{CDM}$ model ($\xi$ free, $w_0 = -1.001$), while Panel (b) corresponds to the $\xi w\text{CDM}$ model ($\xi$ and $w_0$ free). The gray shaded region represents the $1\sigma$ and $2\sigma$ measurement bounds from SH0ES ($H_0 = 73.04 \pm 1.04~\text{km s}^{-1}\text{Mpc}^{-1}$). The addition of external data combinations progressively tightens the $H_0$ constraints while demonstrating a significant reduction in the $H_0$ tension across both interacting scenarios.}
        \label{fig:H0_comparison}
\end{figure}

The physical effect of the negative coupling is a continuous transfer of energy from dark energy to dark matter modifying the background energy density evolution. Because of this continuous energy transfer over cosmic time, fitting the CMB acoustic scale constraints ($\Omega_c h^2$) at early times  demands a modified matter density path. To compensate for the modified late-time expansion history while preserving early-time CMB peak alignments (constraints), the model allows the inferred present-day Hubble constant $H_0$ to shift toward higher values. At the same time, the phantom equation of state $(w_0=-1.116_{-0.043}^{+0.043})$ compensates the continuous loss of energy by increasing the dark energy density as $a \to 1$, giving rise to the required late-time acceleration from it under the perturbation stability condition $\xi(1+w_0)>0$. The combination of these two mechanisms allows the model to produce a value of $H_0 = 70.22_{-0.40}^{+0.45}~\text{km s}^{-1}\text{Mpc}^{-1}$ using the full dataset, resulting in the reduction of the tension with  $\text{SH0ES}$ to $\sim 2.5\sigma$ while generating enhanced fit ($\Delta\chi^2 = -14.60$) over standard $\Lambda\text{CDM}$. Figure~\ref{fig:H0_comparison} demonstrates a direct comparison between the $H_0$  constraints of the two models, with panel (b) emphasizing the broader $H_0$ posterior distribution allowed when both $\xi$ and $w_0$ are free. \\

\noindent The inclusion of DESI DR2 BAO data is specifically effective in breaking such degeneracies. The BAO measurement technique directly constrains the expansion rate $H(z)$ and the angular diameter distance $D_A(z)$ at several redshifts, thus yielding very strong constraints on the late time evolution of the universe. This explains the substantial improvement in the $(w_0, H_0)$ constraints when the BAO data are included (see Figure~\ref{fig:corner_plot2}). The increased precision of DESI DR2 BAO data, compared with the earlier BAO surveys, is therefore significant for obtaining reliable constraints on interacting dark energy models.\\
\noindent As mentioned in Sec.~\ref{sec:constraining power} for the $\xi$CDM model, it is the consistency of geometry from early to late times that drives the reduction of $H_0$ value when BAO and SN are included. Similar mechanism works for the current case, with the extra degeneracy between $\xi$ and $w_0$.

\subsubsection{Dark Energy Equation of State and Phantom Preference}

\noindent The $\xi w$CDM model enables us to determine whether the data require a dynamical dark energy component in addition to coupling. Table~\ref{tab:constraints_xi_w0} shows a consistent preference for a phantom equation of state ($w_0 < -1$) in all dataset combinations. For the Planck-only dataset, we obtain $w_0 = -1.103^{+0.060}_{-0.065}$, which deviates by about $1.7\sigma$ from the cosmological constant. When the complete dataset combination is accounted for, we find $w_0 = -1.116^{+0.043}_{-0.043}$, which lies about 2.7\(\sigma\) below \(w_0 = -1\) within the restricted phantom branch.\\
Given the stability requirement $\xi(1+w_0)>0$ and our choice of $\xi < 0$, this phantom preference is expected.  It is essential to remember that the dataset combination has significant impact on the $w_0$ constraint. When BAO and SNe data are incorporated (Planck+BAO+SN), the central value moves towards $w_0 = -1.049^{+0.029}_{-0.049}$, which is nearer to the cosmological constant. The inclusion of weak lensing data shifts it to $w_0 = -1.116^{+0.043}_{-0.043}$ suggesting some tensions between datasets concerning the preferred equation of state. This behavior is illustrated in figure~\ref{fig:corner_plot2} through two-dimensional contours in the $(w_0, H_0)$ plane.

\subsubsection{Coupling Constraints in the $\xi w$CDM Model}

\begin{figure}[H]
	\centering
	\includegraphics[width=0.50\textwidth]{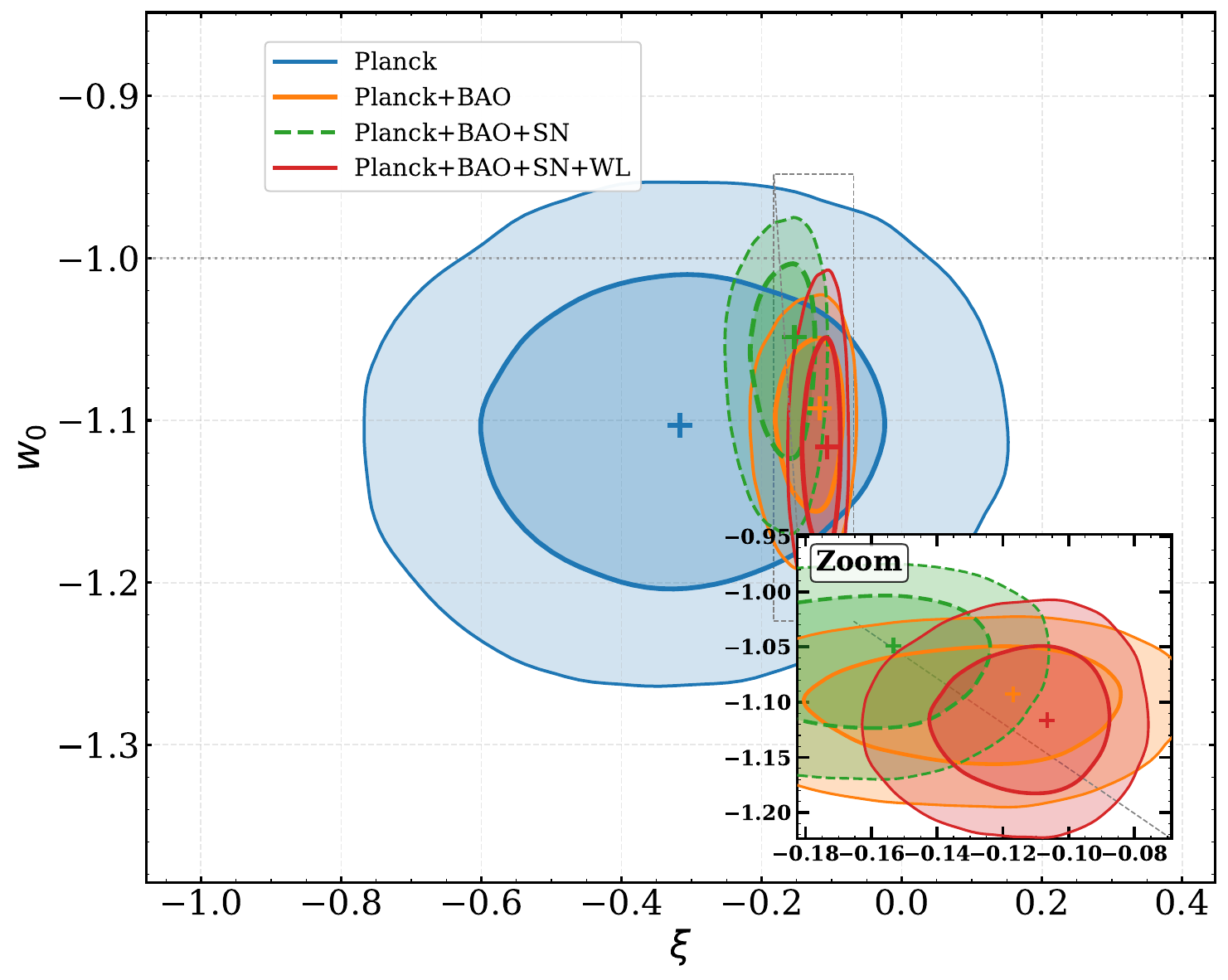}
	\caption{Two-dimensional marginalized posterior distribution (68\% and 95\% confidence levels) for the parameter pair $\xi$-$w_0$, for Planck (blue), Planck+BAO (orange), Planck+BAO+SN (green), and Planck+BAO+SN+WL (red). The '+' markers indicate the best-fit values, and the dashed gray line denotes $w_0 = -1$ (cosmological constant). The inset shows a zoomed region around the combined dataset contours (excluding Planck).}
\label{fig:xi_vs_w0_plot}
\end{figure}

\noindent In the $\xi w$CDM model, the coupling $\xi$ is constrained to be negative in agreement with the stability criterion $\xi(1+w_0)>0$. Due to the considerable degeneracy between $\xi$ and $w_0$ (seen in figure~\ref{fig:xi_vs_w0_plot}),

\noindent we get $\xi = -0.317^{+0.188}_{-0.182}$ for the Planck-only dataset, with significant uncertainty.\\
\noindent Inclusion of the BAO data tightens the constraint to $\xi = -0.117^{+0.021}_{-0.041}$. The complete dataset yields  $\xi = -0.107^{+0.012}_{-0.023}$. When we compare with the $\xi$CDM model,  the $\xi$ constraints are consistent within the uncertainties; however, the $\xi w$CDM model yields more negative values when SNe and WL data are included. \\
\noindent Notably, the Planck + BAO + SN dataset combination provides $\xi = -0.153^{+0.019}_{-0.040}$, which is more negative than the equivalent value in the $\xi$CDM model ($\xi = -0.076^{+0.019}_{-0.030}$). This implies that when the equation of state is allowed to vary, the data prefer a stronger coupling, partly compensated by a less phantom $w_0$. This degeneracy is a prominent feature of the $\xi w$CDM model and is visualized in figure~\ref{fig:xi_vs_w0_plot}.\\
\noindent The inclusion of the KiDS-1000 WL data plays a significant role in the multi-parameter space of the  $\xi w\text{CDM}$ model. This can be seen from Table~\ref{tab:constraints_xi_w0}, where the addition of the SN data yields $H_0 = 71.10^{+0.83}_{-0.45}~\mathrm{km\,s^{-1}\,Mpc^{-1}}$ with a weaker coupling ($\xi = -0.153$). However, the incorporation of the WL data constrains the interaction strength to $\xi = -0.107^{+0.012}_{-0.023}$, and shifts $H_0$ back to $70.22^{+0.45}_{-0.40}~\mathrm{km\,s^{-1}\,Mpc^{-1}}$ and substantially reducing its $68\%$ C.L. uncertainty to about $\pm 0.42~\mathrm{km\,s^{-1}\,Mpc^{-1}}$. At the same time, the inclusion of WL data lowers $\sigma_8$ from $0.8662$ ($\text{Planck}+\text{BAO}+\text{SN}$) to $0.8523$ ($\text{Planck}+\text{BAO}+\text{SN}+\text{WL}$). This shift demonstrates the supporting  role of WL data in constraining the interaction and structure-growth parameters and, as a result, in tightening the allowed cosmological parameter space.\\
\noindent As in the case of $\xi$CDM model, the DESI DR2 BAO data has also been very powerful to break the degeneracies in the $\xi w$CDM model. Figure~\ref{fig:corner_plot2} visualizes this strong constraining power of the DESI DR2 BAO data in breaking the parameter degeneracies in the interacting dark energy model ($\xi w$\text{CDM}). In this model, the constraints are markedly tightened by the inclusion of DESI DR2 BAO measurements compared to Planck-only (blue contours), particularly in the $(\xi, H_0)$, $(w_0, H_0)$, $(H_0, \Omega_m)$, and $(H_0, \Omega_\Lambda)$ planes. \\
\noindent The most remarkable feature is the pronounced tightening of the $(\xi, H_0)$  degeneracy in the presence of DESI DR2 BAO data.  In  $\xi w$CDM model (see Table~\ref{tab:constraints_xi_w0}), an even greater improvement is observed on the $(w_0, H_0)$ plane. The Planck-only data provide $H_0 = 72.46^{+4.05}_{-3.66}$ km s$^{-1}$ Mpc$^{-1}$, with the posterior covering from about $68.8$ to $76.5$ km s$^{-1}$ Mpc$^{-1}$ at 68\% C.L. We obtain $w_0 = -1.103^{+0.060}_{-0.065}$, ranging from approximately $-1.17$ to $-1.04$. When DESI DR2 BAO data are included, the constraints tighten strongly: $H_0 = 70.20^{+0.87}_{-0.52}$ km s$^{-1}$ Mpc$^{-1}$ and $w_0 = -1.093^{+0.027}_{-0.041}$ for the Planck + BAO combination. The full dataset combination yields $H_0 = 70.22^{+0.45}_{-0.40}$ km s$^{-1}$ Mpc$^{-1}$ and $w_0 = -1.116^{+0.043}_{-0.043}$. The uncertainty in $H_0$ becomes four times smaller (from $\sim 7.7$ to $\sim 0.85$ km s$^{-1}$ Mpc$^{-1}$), while the $w_0$ uncertainty improves by a factor of $\sim 1.5$ (from $\sim 0.13$ to $\sim 0.09$). The $(w_0, \xi)$ plane further shows the power of DESI DR2 BAO data. The planck-only contours illustrate a complicated banana-like shape of degeneracy suggesting the compromise between coupling and equation of state. However, these degeneracies are notably resolved when DESI DR2 BAO data are implemented.

\subsubsection{Model Comparison and Parameter Penalty}
\noindent Model comparison statistics indicate a clear picture suggesting that no single model is widely favored across all dataset combinations. For dataset combination Planck + BAO, we obtain  $\Delta\mathrm{AIC} = +3.31$ and $\Delta\mathrm{BIC} = +5.20$,  indicating that $\Lambda$CDM model is preferred when only BAO data are included. However, this is to be interpreted as a conservative estimate, since BAO data mainly constrain the expansion history and may not be sensitive to all the possible effects of interactions in the dark sector.\\
Adding the SN data (Planck + BAO + SN), improves the fit significantly ($\Delta\chi^2 = -14.60$) with $\Delta\mathrm{AIC} = -10.60$, indicating that the fit improvement strongly favors the $\xi w$CDM model according to the AIC scale. Although $\Delta\mathrm{BIC}=+0.30$ shows a weak preference for  $\Lambda\mathrm{CDM}$ due to the parameter penalty, the magnitude $|\Delta\mathrm{BIC}| < 2$ indicates that this preference is statistically negligible. The small difference in BIC values ($|\Delta\mathrm{BIC}| < 2$) implies that the $\xi w$CDM model, although with two more parameters, is statistically competitive with $\Lambda$CDM based on the conservative Bayesian criterion. A modest improvement in the quality of the data could thus shift the preference to the interacting model.\\ 
For the full dataset combination (Planck + BAO + SN + WL), we find $\Delta\mathrm{AIC} = -2.67$, which according to the AIC scale shows a \emph{positive but weak} preference for the $\xi w$CDM model. However, $\Delta\mathrm{BIC} = +8.23$ strongly favors $\Lambda$CDM. The BIC value of $+8.23$ is above the threshold of $6$, indicating \emph{strong evidence} against the $\xi w$CDM model in favor of $\Lambda$CDM according to the standard thresholds. The fact that the AIC still prefers the interacting model even when including WL data indicates that the $\xi w$CDM model contains significant physical aspects that are worth studying in more detail.\\
The persistent preference for negative $\xi$ and phantom $w_0$, the large improvement in fit upon inclusion of SNe data ($\Delta\chi^2 = -14.60$), together with the near-degeneracy of the BIC for the Planck+BAO+SN combination \((\Delta\mathrm{BIC} = +0.30)\), point to interacting dark energy as a phenomenologically viable extension. For the full Planck+BAO+SN+WL combination, however, the BIC strongly penalizes the two additional parameters \((\Delta\mathrm{BIC} = +8.23)\), so the statistical preference for the interacting model is dataset-dependent. 
\section{Conclusion}
\label{sec:conc}
\noindent In this study, we have carried out a complete analysis of two IDE models, namely the $\xi$CDM and $\xi w$CDM models, incorporating the latest DESI DR2 BAO, Planck 2018 CMB, SN(pantheon+) and KiDS-1000/DES-Y3 weak lensing data. Our findings demonstrate that dark sector interactions allow a motivated way toward reducing unresolved Hubble tension. \\
For the $\xi$CDM model, in which the coupling parameter $\xi$ is free while the DE equation of state $w_0=-1.001$ is fixed, we get $H_0 = 71.59^{+1.69}_{-2.30}\ \mathrm{km\,s^{-1}\,Mpc^{-1}}$ from Planck-only data, tightening to $H_0 = 69.72^{+0.50}_{-0.41}\ \mathrm{km\,s^{-1}\,Mpc^{-1}}$ with the full dataset. This shows a significant reduction of the Hubble tension from $\sim 5.0\sigma$ to $\sim 2.8\sigma$, equivalent to a relative decrease of $44\%$. The coupling parameter is constrained to be negative ($\xi \approx -0.074$ with the full dataset), demonstrating energy transfer from dark energy to dark matter, which naturally requires a higher $H_0$ to remain consistent with the CMB constraint on the physical cold-dark-matter density \(\Omega_c h^2\). Model comparison provides a fit enhancement of $\Delta\chi^2 = -7.00$ ($\text{Planck}+\text{BAO}+\text{SN}+\text{WL}$) with AIC showing strong preference ($\Delta\mathrm{AIC} = -5.00$) and BIC persists weak $\Lambda$CDM preference ($\Delta\mathrm{BIC} = +0.45$). \\
The $\xi w$CDM model, where both $\xi$ and $w_0$ are free parameters meeting the stability criterion $\xi(1+w_0)>0$, provides even stronger predictions. For the Planck only-data, we obtain $H_0 = 72.46^{+4.05}_{-3.66}\ \mathrm{km\,s^{-1}\,Mpc^{-1}}$, improving to  $H_0 = 71.10^{+0.83}_{-0.45}\ \mathrm{km\,s^{-1}\,Mpc^{-1}}$ for Planck+BAO+SN, and $H_0 = 70.22^{+0.45}_{-0.40}\ \mathrm{km\,s^{-1}\,Mpc^{-1}}$ for the full dataset, reducing the tension to $\sim 2.5\sigma$. The dark energy equation of state indicates a stable preference for the phantom regime with $w_0 = -1.116^{+0.043}_{-0.043}$ from the complete dataset, corresponding to a $2.7\sigma$ deviations from the cosmological-constant value ($w_0=-1$). Particularly, the fit obtained for the $\xi w$CDM model using the Planck+BAO+SN compilation indicates a substantial  improvement in fit ($\Delta\chi^2=-14.60$) together with a remarkable AIC  preference ($\Delta\mathrm{AIC}=-10.60$). Although the conservative BIC criterion indicates a weak preference for \(\Lambda\)CDM for the Planck+BAO+SN combination \((\Delta\mathrm{BIC} = +0.30)\) and a strong preference for the full combination \((\Delta\mathrm{BIC} = +8.23)\) due to the parameter penalty, the combination of strong preference for AIC, improved fit, and near degeneracy in BIC suggests that the model $\xi w$CDM is statistically credible compared to the $\Lambda$CDM model.\\
Our result analysis also stresses on the highly constraining nature of the DESI DR2 BAO data, that improve $H_0$ constraints by a factor of 4-8, while breaking essential degeneracies between  $\xi$, $w_0$, and $H_0$, degeneracies that have been challenging for investigating dark sector interactions. The gradual tightening from Planck-only data to full dataset together indicates the combined aspects of these probes and their importance in the proper constraining of IDE models.\\
The persistence of the negative coupling and phantom equation of state preference, the significant improvement of the fit when SNe data is included and the near degeneracy in Bayesian comparison all point to interacting dark energy as a physically motivated mechanism for alleviating the $H_0$ tension. Our results show that dark sector interactions yield a viable physical process for uniting early and late universe expansion rate measurements, but an alleviation of the Hubble tension will likely need more sophisticated phenomenological extensions or next generation cosmological surveys. 


\bibliographystyle{elsarticle-num}
\bibliography{references_DE}

\end{document}